\documentclass[pdflatex]{article}
\usepackage{graphicx}
\usepackage{subfigure}
\usepackage{wrapfig}
\usepackage{amsmath}
\begin{document} 
\title{Primordial black hole formation from
cosmological fluctuations
}

 \author{Tomohiro Harada\\
 {\small  Department of Physics, Rikkyo University, Toshima, Tokyo 171-8501, Japan} \\
 {\small {\em Email} : harada@rikkyo.ac.jp}}

%
\maketitle
\begin{abstract} 
Primordial black holes (PBHs) are those which may have formed in the
 early Universe and affected the subsequent evolution of the Universe
 through their Hawking radiation and gravitational field. To constrain
 the early Universe from the observational constraint on 
 the abundance of PBHs, it is essential to determine the formation
 threshold for primordial cosmological fluctuations, which are naturally
 described by cosmological long-wavelength solutions. 
I will briefly review our recent analytical and numerical results 
on the PBH formation.
\end{abstract} 
\section{Introduction} 

Recently, primordial black holes (PBHs) have been intensively studied
because of their unique role in cosmology.
PBHs have affected our Universe through their Hawking radiation, gravitational
force and gravitational radiation. Because of this nature of PBHs,  
the current observational data can strongly constrain the abundance of PBHs. 
Since the PBHs of mass $M$ are formed at the epoch when the mass contained 
within the Hubble horizon is $M$, 
we can put constraints on the early Universe through PBHs. 
Therefore, we can regard PBHs as the fossils of the early Universe.
The quantitative studies on the early Universe through the observational
constraint on the abundance of PBHs have been pioneered 
by Carr~\cite{Carr:1975qj} and recently updated 
by Carr et al.~\cite{Carr:2009jm}.

The PBHs of mass $M$ are formed from the density fluctuations of 
mass $M$, which we denote as $\delta(M)$.
If we assume Gaussian probability distribution for $\delta(M)$,
though this assumption must be modified in the nonlinear 
regime~\cite{Kopp:2010sh},
the production rate of the PBHs is given by 
\begin{eqnarray*}
\beta_{0}(M)&\simeq& \sqrt{\frac{2}{\pi}}\frac{\sigma(M)}{\delta_{c}(M)}
 \exp\left(-\frac{\delta_{c}^{2}(M)}{2\sigma^{2}(M)}\right), 
\end{eqnarray*}
where $\delta_{c}(M)$ is the threshold 
and $\sigma(M)$ is the standard deviation of $\delta(M)$.
Note that $\delta_{c}$ is of order unity in the phase when the 
Universe is dominated by a relativistic fluid, e.g., 
a radiation fluid.  

\section{Analytic threshold formula}

We focus on the scenario in which PBHs are formed from 
fluctuations generated by inflation. 
Inflation can generate fluctuations of long-wavelength scale, i.e., $L\gg
H^{-1}$, where $H$ is the Hubble parameter and we use the units in
which $c=G=1$.
After the inflationary phase, the Universe undergoes 
decelerated expansion. When 
the fluctuations of scale $L$ enter the Hubble horizon, where 
$L=H^{-1}$, 
the mass scale $M$ of the fluctuation is equal to the mass $M_{H}$ 
      contained within the Hubble radius.
If the amplitude of the fluctuation is sufficiently large, 
the Jeans instability sets in and the 
fluctuation begins to collapse. Since 
the gravitational radius of $M_{H}$ is equal to the Hubble
      radius, a black hole apparent 
      horizon is formed soon after the collapse begins.

\begin{wrapfigure}[14]{r}{5cm}
\begin{center}
\includegraphics[width=0.35\textwidth]{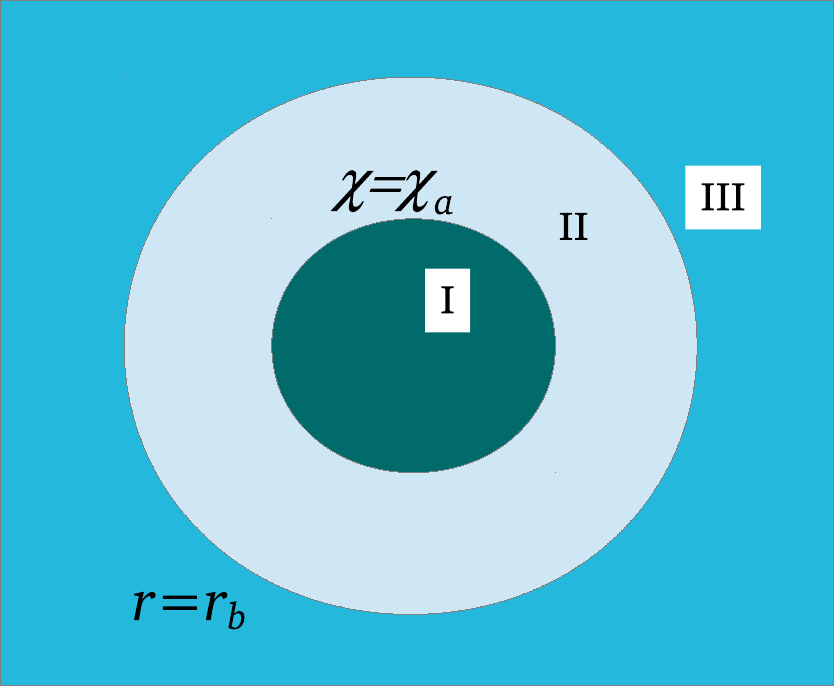}
\end{center}
\caption{\label{fg:density_perturbation_model} The 3-zone model.}
\end{wrapfigure}

First, we discuss an analytic approach to the determination of the 
black hole threshold. 
We adopt the following model, which we call the 3-zone model. 
See Fig.~\ref{fg:density_perturbation_model}.
We assume a flat Friedmann-Lemaitre-Robertson-Walker (FLRW) 
universe surrounding the 
overdense region (region III). 
The line element in the flat FLRW spacetime
is applicable only for $r>r_{b}$, where $r$ is the comoving radial
coordinate in the flat space.
The central region (region I) 
undergoes expansion, maximum expansion and collapse
to singularity. It is described by the closed FLRW
spacetime.
The line element in the closed FLRW spacetime 
is applicable only for $0\le \chi<\chi_{a}$, where $\chi$ is the
comoving radial coordinate in the unit 3-sphere.
We need a compensating layer (region II) between regions I and III. 

The Jeans criterion tells us that the pressure gradient force
cannot suppress gravitational instability if and only if 
the free-fall time $t_{\rm ff}$
is shorter than 
the sound-crossing time $t_{\rm sc}$.
Since we are concerned with a highly general relativistic system,
we need to be precise for the application of the Jeans criterion.
Here, we adopt the following criterion: 
if and only if the overdense region ends in singularity
before a sound wave crosses its diameter from the big bang, 
it collapses to a black hole.

We assume an equation of state (EOS) $p=(\Gamma-1)\rho$ for simplicity. 
For example, $\Gamma=4/3$ corresponds to a radiation fluid.
We define
$\tilde{\delta}$ as the density perturbation at horizon entry 
in the comoving slicing. 
Carr~\cite{Carr:1975qj} proposed a formula $\tilde{\delta}_{{\rm CMC},
c}\simeq \Gamma-1$ and $\tilde{\delta}_{{\rm CMC}, {\rm max}}\simeq 1$ 
in the constant-mean-curvature (CMC) slicing
based on some Newtonian-like approximation.
The latter is the maximum value that the perturbation can 
take at the horizon entry. 
This is transformed in the comoving slicing to 
$
\tilde{\delta}_{c}= 3\Gamma (\Gamma-1)/(3\Gamma+2),~~
 \tilde{\delta}_{\rm max}=3\Gamma/(3\Gamma+2)$.
We can refine Carr's formula to 
the following~\cite{Harada:2013}:
\begin{equation}
 \tilde{\delta}_{c}=\frac{3\Gamma}{3\Gamma+2}
\sin^{2}\left(\frac{\pi\sqrt{\Gamma-1}}{3\Gamma-2}\right),
\quad 
 \tilde{\delta}_{\rm max}=\frac{3\Gamma}{3\Gamma+2}.
\label{eq:HYK_formula}
\end{equation}
This new formula does not rely on any Newtonian-like approximation.
The $\sin^{2}$ dependence comes from the 
spherical geometry of the overdense region and 
$(3\Gamma-2)$ in the denominator in the argument of $\sin^{2}$ 
comes from the fact that the pressure $p$ as well as the density $\rho$
appears as the source of the gravitational field in the form $(\rho+3p)$.

\section{Primordial fluctuations}

\begin{wrapfigure}[12]{r}{5cm}
\begin{center}
 \includegraphics[width=0.4\textwidth]{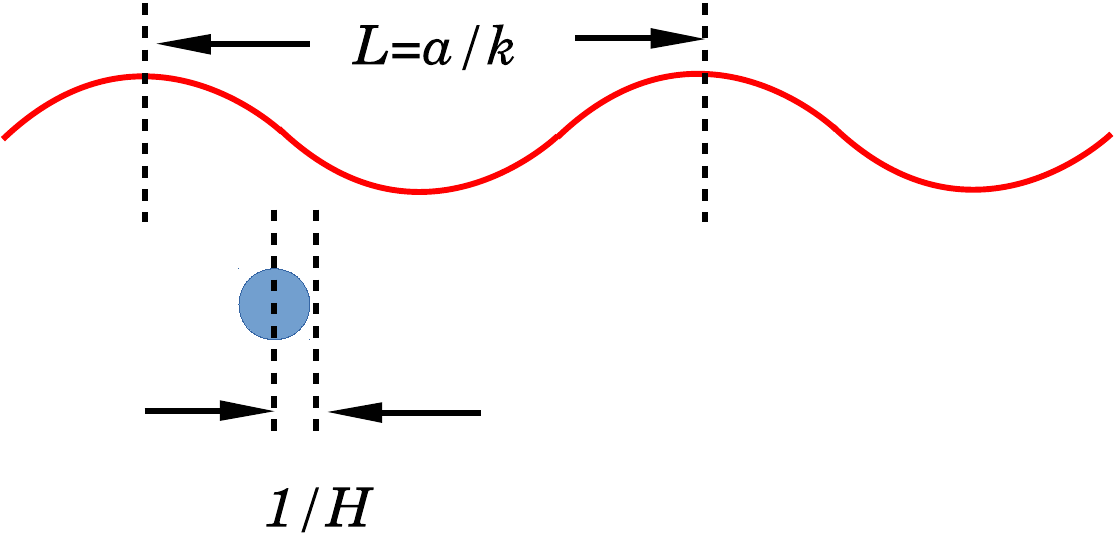}
\end{center}
\caption{\label{fg:long_wavelength} The CLWL solutions.}
\end{wrapfigure}

Although the 3-zone model is powerful in deriving the new analytic formula,
it is by no means generic.
It would be important to construct more general cosmological
fluctuations. This can be done by cosmological 
long-wavelength (CLWL) solutions. Here, the spacetime is assumed to be smooth 
in the scales larger than $L=a/k$, which is much longer than the local 
Hubble length $H^{-1}$. Under this assumption, 
      by gradient expansion, the exact solution is 
      expanded in powers of $\epsilon \sim k/(aH)$.
The CLWL solutions are schematically
depicted in Fig.~\ref{fg:long_wavelength}.

We decompose the spacetime metric in the following form:
\begin{equation}
 ds^{2}=-\alpha^{2}dt^{2}+\psi^{4}a^{2}(t)\tilde{\gamma}_{ij}(dx^{i}+\beta^{i}dt)(dx^{j}+\beta^{j}dt),
\end{equation}
and we choose $\tilde{\gamma}=\eta$, where
$\tilde{\gamma}=\mbox{det}(\tilde{\gamma}_{ij})$,
$\eta=\mbox{det}(\eta_{ij})$, and $\eta_{ij}$ is the metric of the
      3 dimensional flat space.
We call the above decomposition the cosmological conformal decomposition.
According to Lyth et al.~\cite{Lyth:2004gb}, we assume the spacetime
locally approaches 
the flat FLRW form in the limit $\epsilon\to 0$.
Then, the Einstein equations in $O(1)$ imply the Friedmann equation
and the energy equation.
We can deduce $\psi=\Psi(x^{i})+O(\epsilon^{2})$ for
a perfect fluid with a barotropic EOS. $\Psi(x^{i})$ generates the
solution. Thus, the different slicings are equivalent to $O(\epsilon)$.

In the following, we focus on spherically symmetric systems.
Two independent approaches have been adopted for simulating PBH formation from
cosmological fluctuations.
The one is based on the CMC slicing and conformally flat spatial
coordinates.
This is adopted by Shibata and Sasaki~\cite{Shibata:1999zs}.
We denote the radial coordinate in this scheme with $\varpi$.
The initial conditions are given by the CLWL solutions generated
by $\Psi(\varpi)$, where $\psi=\Psi(\varpi)+O(\epsilon^{2})$. 
The other is based on the comoving slicing and comoving threading.
We denote the radial coordinate in this scheme with $r$.
Polnarev and Musco~\cite{Polnarev:2006aa} constructed the initial
conditions as follows: the metric is assumed to approach
\begin{equation}
 ds^{2}=-dt^{2}+a^{2}(t)\left[\frac{dr^{2}}{1-K(r)r^{2}}+r^{2}(d\theta^{2}+\sin^{2}\theta
      d\phi^{2})\right],
\end{equation}
in the limit $\epsilon\to 0$. The exact solution is expanded in
      powers of $\epsilon$ and generated by $K(r)$.
These solutions are called asymptotically quasi-homogeneous solutions.
In fact, we can show that 
the CLWL solutions and the Polnarev-Musco 
asymptotic quasi-homogeneous solutions 
are equivalent with each other through the relation~\cite{Harada:2015}
\begin{eqnarray}
r=\Psi^{2}(\varpi)\varpi,\quad
K(r)r^{2}=1-\left(1+2\displaystyle\frac{\varpi}{\Psi(\varpi)} \frac{d\Psi(\varpi)}{d\varpi}\right)^{2}.
\end{eqnarray}
We can also derive the correspondence relations between the two solutions.

\section{Numerical results}

\begin{wrapfigure}[17]{r}{7cm}
\begin{center}
 \includegraphics[width=0.5\textwidth]{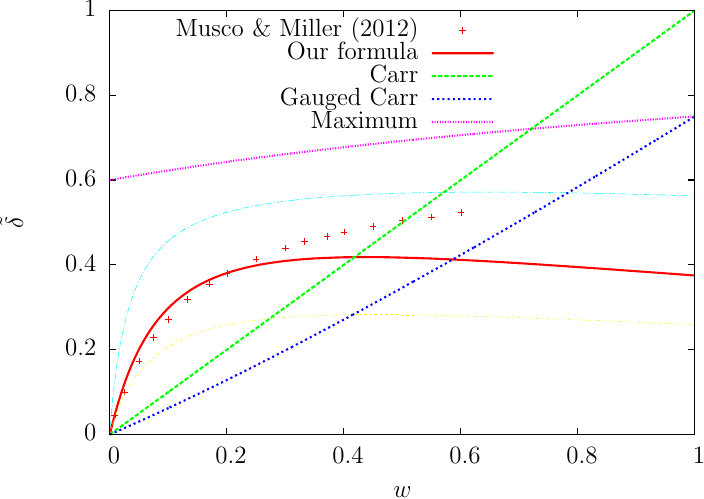}
\caption{\label{fg:eos_dependence} Carr's formula and 
the new formula are plotted together with the 
numerical result~\cite{Musco:2012au}
for $0.01\le w \le 0.6$, where $w=\Gamma-1$.}
\end{center}
\end{wrapfigure}

It is useful to define the amplitude of perturbation for the description 
of the black hole threshold.
We define $\tilde{\delta}$ as 
the density perturbation at the horizon entry by 
$
\tilde{\delta}:=\lim_{\epsilon\to
      0}\bar{\delta}_{\rm C}(t, r_{0})\epsilon^{-2},
$
where $\bar{\delta}_{\rm C}(t, r_{0})$ is the density in the comoving
      slicing averaged over 
$r_{0}$, the radius of the overdense region.
$\tilde{\delta}$ can be directly calculated from $\Psi(\varpi)$ or $K(r)$.
$\psi_{0}$ is the initial peak value of the curvature variable 
$
\psi_{0}:=\Psi(0)$,
which is another independent amplitude of perturbation.
Thus, we can calculate the threshold value of the amplitude
after we have determined the black hole threshold by 
numerical simulations. 
See~\cite{Harada:2015} for the details of the numerical simulations.
Our numerical results are consistent and complementary to 
Refs.~\cite{Shibata:1999zs,Polnarev:2006aa,Musco:2012au}. 
We will present and interpret the numerical results below. 

First we see the dependence of the threshold $\tilde{\delta}_{c}$ on the EOS.
In Fig.~\ref{fg:eos_dependence}, we plot Carr's
formula, the new formula~(\ref{eq:HYK_formula})
together with the numerical result for a Gaussian curvature profile 
in Musco and Miller~\cite{Musco:2012au}.
We can see that Carr's formula underestimates the threshold by 
a factor of two to ten, while the new formula agrees with the 
numerical result within 20 percent, although 
both formulas are consistent with the numerical result 
in the order of magnitude.

\begin{figure}[htbp]
\begin{center}
\subfigure[]{\includegraphics[width=0.4\textwidth]{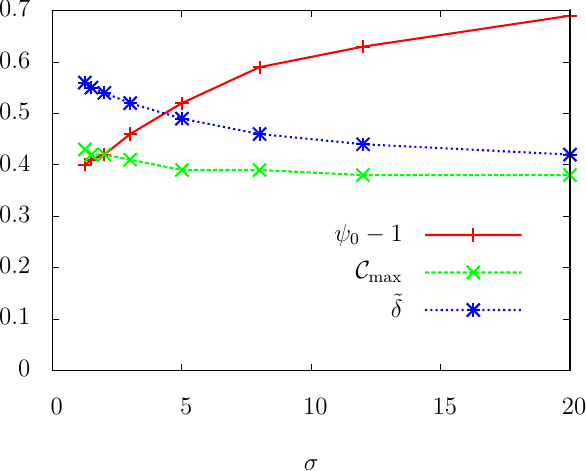}
 \label{fg:SS99_initial_data}}
\subfigure[]{\includegraphics[width=0.4\textwidth]{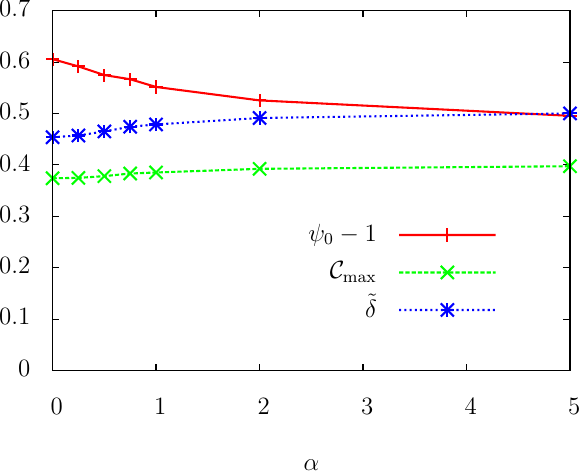}
 \label{fig:PM07_initial_data}}
\caption{\label{fg:threshold}
The threshold values $\tilde{\delta}_{c}$ and $\psi_{0,c}$ are 
plotted for the Gaussian-type initial 
profiles in the parametrisations~in \cite{Shibata:1999zs} 
(a) and 
in~\cite{Polnarev:2006aa} (b).}  
\end{center}
\end{figure}

Next we see the dependence of the threshold on the initial profile.
Shibata and Sasaki~\cite{Shibata:1999zs} and Polnarev and
Musco~\cite{Polnarev:2006aa} choose  
similar Gaussian-type profiles but with different parametrisations. 
In the former, the profile is parametrised by $\sigma$, in which the
smaller the $\sigma$ is, the sharper the transition from the overdense
region to the surrounding flat FLRW region becomes.
In the latter, it is parametrised by $\alpha$, in which the larger the
$\alpha$ is, the sharper the transition becomes.
Our numerical results are plotted in Fig.~\ref{fg:threshold}
for $\tilde{\delta}_{c}$ and $\psi_{0,c}$.
As we can see, the sharper the transition is, 
the larger the $\tilde{\delta}_{c}$ is but the smaller 
the $\psi_{0,c}$ is.

We can naturally interpret the behaviour of $\tilde{\delta}_{c}$. 
To overcome larger pressure gradient force, we need stronger
gravitational force and hence larger amplitude of perturbation.
We can also see that the minimum value of $\tilde{\delta}_{c}$,
which is realised in the gentlest transition, 
is close to the value given by the new 
       formula, while the maximum value, which 
is realised in the sharpest transition, 
is close to the maximum value given by the new formula.
That is,
$
 \tilde{\delta}_{c, {\rm
  min}}<\tilde{\delta}_{c}<\tilde{\delta}_{c, {\rm max}}$, 
where 
\begin{equation}
 \tilde{\delta}_{c, {\rm
       min}}\simeq \frac{3\Gamma}{3\Gamma+2}\sin^{2}\left(\frac{\pi\sqrt{\Gamma-1}}{3\Gamma-2}\right),\quad 
 \tilde{\delta}_{c, {\rm
  max}}\simeq \frac{3\Gamma}{3\Gamma+2}.
\label{eq:min_max}
\end{equation}

As for $\psi_{0,c}$, the behaviour is apparently opposite to that of  
$\tilde{\delta}_{c}$. The sharper the transition is, the smaller the  
$\psi_{0,c}$ is. We can interpret this by the nonlocal nature of $\psi$.
In fact, $\psi$ is affected by the perturbation 
in the far region, which spreads even outside the local Hubble 
horizon, while the threshold should be determined by quasi-local dynamics 
within the local Hubble length. 
$\psi_{0,c}$ is sensitive to the
environment, while 
$\tilde{\delta}_{c}$ is not.

\section{Conclusion}

We have seen that the Jeans criterion 
applied to the simplified toy model gives a new analytic formula for the
threshold of PBH formation.
For more general situations, the CLWL solutions 
naturally give primordial fluctuations generated by inflation.
We have numerically found that 
the sharper the transition from the overdense region to the surrounding
flat FLRW universe is, the larger the density perturbation $\tilde{\delta}_{c}$ 
at the threshold becomes.
The new formula is supported by the numerical result in the form given
by Eq.~(\ref{eq:min_max}). 
The peak value of the curvature variable $\psi_{0,c}$ is 
subjected to the environmental effect and hence will not provide a 
compact criterion for the PBH formation.

\vspace{0.2cm}
This article is based on the
collaboration~\cite{Harada:2013,Harada:2015} 
with C.~M.~Yoo, K.~Kohri,
T.~Nakama and Y.~Koga.
This work was supported by JSPS KAKENHI Grant Number 26400282.


\begin{thebibliography}{00}

\bibitem{Carr:1975qj}
  B.~J.~Carr,
  ``The Primordial black hole mass spectrum,''
  Astrophys.\ J.\  {\bf 201}, 1 (1975).
\bibitem{Carr:2009jm}
  B.~J.~Carr, K.~Kohri, Y.~Sendouda, and J.~Yokoyama,
  ``New cosmological constraints on primordial black holes,''
  Phys.\ Rev.\ D {\bf 81}, 104019 (2010).
  [arXiv:0912.5297 [astro-ph.CO]].
\bibitem{Kopp:2010sh}
  M.~Kopp, S.~Hofmann and J.~Weller,
  ``Separate Universes Do Not Constrain Primordial Black Hole Formation,''
  Phys.\ Rev.\ D {\bf 83} (2011) 124025
  [arXiv:1012.4369 [astro-ph.CO]].
\bibitem{Harada:2013} T. Harada, C.M. Yoo and K. Kohri, Phys. Rev. D 88 (2013) 8, 084051
 [Phys. Rev. D 89 (2014) 2, 029903] [arXiv:1309.4201 [astro-ph.CO]].
\bibitem{Lyth:2004gb}
  D.~H.~Lyth, K.~A.~Malik and M.~Sasaki,
  ``A general proof of the conservation of the curvature perturbation,''
  JCAP {\bf 05} (2005) 004
  [astro-ph/0411220].
\bibitem{Shibata:1999zs}
  M.~Shibata and M.~Sasaki,
  ``Black hole formation in the Friedmann universe: Formulation and computation in numerical relativity,''
  Phys.\ Rev.\ D {\bf 60} (1999) 084002
  [gr-qc/9905064].
\bibitem{Polnarev:2006aa}
  A.~G.~Polnarev and I.~Musco,
  ``Curvature profiles as initial conditions for primordial black hole formation,''
  Class.\ Quant.\ Grav.\  {\bf 24} (2007) 1405
  [gr-qc/0605122].
\bibitem{Musco:2012au}
  I.~Musco and J.~C.~Miller,
  ``Primordial black hole formation in the early universe: critical behaviour and self-similarity,''
  Class.\ Quant.\ Grav.\  {\bf 30} (2013) 145009
  [arXiv:1201.2379 [gr-qc]].
\bibitem{Harada:2015} T. Harada, C.M. Yoo, T.~Nakama and Y. Koga, Phys. Rev. D 91 (2015) 8, 084057 [arXiv:1503.03934 [gr-qc]].
\end{thebibliography}
\end{document}